\documentstyle[eqsecnum,floats,aps,prb,epsfig,twocolumn]{revtex}

\begin{document}

\wideabs{

\title { Microscopic theory of quadrupolar ordering in TmTe } 
 
\author { 
A.V.~Nikolaev\cite{addr}  and K.H.~Michel \\} 
\address{Department of Physics, University of Antwerp, UIA,\\ 
2610 Antwerpen, Belgium
} 
 
\date{\today} 
 
\maketitle 
 
%--------------- ABSTRACT --------------- 
\begin{abstract}
We have calculated the crystal electric field of TmTe ($T>T_Q$)
and have obtained that the ground state of a Tm 4$f$ hole is the
$\Gamma_7$ doublet in agreement with M\"{o}ssbauer experiments.
We study the quadrupole interactions
arising from quantum transitions of 4$f$ holes of Tm.
An effective attraction is found at the $L$
point of the Brillouin zone, $\vec{q}_L$. Assuming that the
quadrupolar condensation involves a single arm of $\vec{q}_L$ 
we show that there are 
two variants for quadrupole ordering which are described
by the space groups $C2/c$ and $C2/m$.
The Landau free energy is derived in mean-field theory.
The phase transition is of second order.
The corresponding quadrupole order parameters are 
combinations of $T_{2g}$ and $E_g$ components. 
The obtained domain structure is in agreement with
observations from neutron diffraction studies for TmTe.
Calculated lattice distortions are found to be different
for the two variants of quadrupole ordering.
We suggest to
measure lattice displacements in order to discriminate
between those two structures. 
\end{abstract} 
 
% PACS (for EPJ B) are
% 71.10.+x {quantum quadrupolar interaction},
% 71.27.+a {strongly correlated 4f electron system}, 
% 71.45.-d {cooperative 4f-phenomena},
% 64.70.Kb {solid-solid phase transitions in Ce}.
%

}

\narrowtext

\section {Introduction} 
\label{sec:int} 

The monochalcogenides of rare-earth elements
crystallizing in the rocksalt structure,
exhibit interesting electrical, optical and
magnetic properties. \cite{Wac}
In TmTe thulium is in a divalent
state with one 4$f$-hole (4$f^{13}$)
localized at each Tm site.
At room temperature TmTe is a semiconductor
with the electron band energy gap  
0.25 eV (from photoemission data) and an
$f\rightarrow d$ band gap 0.35 eV (from absorption data) \cite{Sur}.
Magnetic structure determination leads to a type II
antiferromagnet below the N\'{e}el temperature \cite{Las} $T_N=0.43$~K. 
Under pressure of 2 GPa TmTe undergoes a 
semiconductor-metal transition \cite{Mat1}
where a ferromagnetic order occurs \cite{Lin} with a Curie
temperature $T_C=14$~K for 2.7 GPa. 
At pressure of 5.7 GPa TmTe undergoes a structural
phase transition to a tetragonal structure \cite{Hea,Tan}
with a further decrease of resistivity. \cite{Mat1}
A renewed interest to TmTe has emerged from the
recent unexpected discovery of an antiferroquadrupolar (AFQ)
phase transition below $T_Q=1.8$~K
at atmospheric pressure. \cite{Mat2}
The transition was first revealed by specific heat measurements \cite{Mat2}
and then confirmed by elastic constant anomalies \cite{Mat3} and
by neutron diffraction in an external magnetic field. \cite{Lin2}
In the quadrupolar ordered state by the application
of a uniform magnetic field superlattice neutron
Bragg peaks, corresponding to the wave vector
$\vec{q}=(2\pi/a)(1/2,1/2,1/2)$, appear in addition to the normal
ferromagnetic intensity superimposed on the nuclear peaks \cite{Lin2}
(see for a review Ref. \onlinecite{Mig}
and for recent development Refs. \onlinecite{Lin3,Mig2}).
Thus TmTe has been found to belong to an increasing family
of rare-earths \cite{Gig} (TmCd, TmAu$_2$, 
CeB$_6$, DyB$_2$C$_2$ and so on)
which exhibit quadrupolar ordering.

An important insight in the nature of the AFQ ordering of TmTe
can be gained on the basis of a classical description
of multipoles presented in Refs.~\onlinecite{Shi1,Shi2}.
However, the theoretical treatment in Refs.~\onlinecite{Shi1,Shi2}
is based on a phenomenological Hamiltonian where the
quadrupole-quadrupole interactions are taken to be isotropic
in the subspaces of $\Gamma_3$ ($E_g$) and $\Gamma_5$ ($T_{2g}$)
irreducible representations of the cubic point group $O_h$.
In addition, a coupling between these representations is
neglected. Such a treatment has shortcomings and does not allow
to determine the crystal structure.
Indeed, since the quadrupolar interaction forces are of
short range, it is necessary to take into account the
structure of the face centered cubic (fcc) lattice and the
full matrix character of the quadrupolar interactions.
Furthermore, the coupling of $T_{2g}$ and $E_g$ representations
is important and affects the quadrupolar phase transition.
It is the purpose of the present paper to investigate the quadrupolar 
phase transition and to determine the crystal structure
(space group) of the quadrupolar ordered phase.
Hence it is necessary that the symmetries of quadrupolar interactions
on a fcc lattice are carefully taken into account.
Multipolar interactions of $E_g$ and $T_{2g}$ modes on a fcc
lattice have been studied earlier in connections with
the problem of orientational order in a molecular crystal, 
solid C$_{60}$. \cite{Mic1} 

In constructing a microscopic theory of quadrupolar ordering in TmTe
we will use the concepts we have developed recently \cite{Nik1,Nik2}
for the description of the $\gamma-\alpha$ transition in Ce.
We will take advantage of the fact that Tm and Ce are mirror elements
in the series of lanthanides. We recall that Ce has one single electron
in the 4$f$ shell while Tm has one hole (electron configuration 4$f^{13}$).
While Ce is a metal, TmTe is an insulator for $T$ near $T_Q$.
Hence we will neglect the polarization of conduction electrons
(which contributes to the quadrupolar interaction \cite{Nik2} in Ce)
in our treatment of TmTe. Because of charge transfer
from Tm to Te, only a small fraction of conduction electron density is
left around the Tm$^{++}$ ion.

In zero magnetic field the magnetic susceptibility of TmTe shows
a small anomaly at $T_Q$. \cite{Mat3}
Thus the quadrupolar ordering is not directly connected \cite{Mat3}
with the magnetic ordering in TmTe which occurs at still lower
temperature $T_N$. Therefore in this paper we limit ourselves to the  
charge degrees of freedom leaving the magnetic properties for
future considerations.

The content of the present paper is the following.
In Section II we calculate the crystal field which refers to the 4$f$
hole in TmTe. The result is compared with experimental data.
Next (Sect. III) we study the quadrupolar interactions and the
resulting phase transition. We suggest condensation schemes
leading to the monoclinic space groups $C2/c$ or $C2/m$.
In order to discriminate between these space groups, we calculate 
the accompanying lattice distortions (Sect. IV) and suggest
synchrotron radiation experiments.
Finally, the results of our work are summarized and commented in
the Conclusions (V).

%----------------- II ----------------------> 
\section {Crystal field} 
\label{sec:cf} 

The issue of the crystal electric field (CEF) of TmTe is controversial.
A Schottky specific heat peak around 5~K has been found
indicating that the total CEF splitting is about 15~K. \cite{Buc}
From thermal expansion data  at low temperature (2-16~K)
the following sequence $\Gamma_8(0$~K$)-\Gamma_7(10$~K$)-\Gamma_6(16$~K)
has been proposed. \cite{Ott}
Recent inelastic neutron scattering experiment supports the
$\Gamma_8$ ground state. \cite{Cle}
However, from M\"{o}ssbauer spectroscopy another scheme,
$\Gamma_7(0$~K$)-\Gamma_8(12$~K$)-\Gamma_6(19.6$~K),
has been deduced. \cite{Tri} The latter sequence was also obtained
by detailed ultrasonic velocity measurements. \cite{Kas}
Here we would like to remark that some of these results are not direct
and depend
on conditions, methods and models used to fit experimental data.

As a starting point of our derivation of the CEF we recall that here we
consider the effect on the hole in the 4$f$ shell of Tm$^{++}$.
In comparison with the case of one electron (as in Ce), the position
of the corresponding energy levels is reversed.
For ions of lanthanides in solids the spin-orbit coupling
\begin{eqnarray}
 V_{so} =  \zeta \vec{L} \cdot \vec{S}
  \label{2.1}  
\end{eqnarray}
dominates crystal field effects. The potential $V_{so}$ has spherical
symmetry. Here $\vec{L}$ and $\vec{S}$ are the orbital and the spin 
angular momentum operators, $\vec{J}=\vec{L}+\vec{S}$ is the total
angular momentum, with eigenvalues of $J_z$: $J=7/2$ and $J=5/2$.
In Eq.~(\ref{2.1}) $\zeta$ is the spin-orbit coupling constant.
In case of a hole, the lower energy level corresponds to the state $D_{7/2}$
(degeneracy 8) and the higher level to $D_{5/2}$ (degeneracy 6).
The experimental energy separation  \cite{Sur} $\triangle_{so}=-(7/2)\zeta$
is given by 1.24 eV, corresponding to $\zeta=-4112$~K.
In presence of the cubic crystal field $V_{CF}$, the hole experiences
the potential
\begin{eqnarray}
 V_{0}^f =  V_{so}+V_{CF} .
  \label{2.2}  
\end{eqnarray}
Since $V_{CF}$ has cubic symmetry, the degeneracies of the spin-orbit levels
are lifted according to the scheme \cite{Lea}
\begin{mathletters}
\begin{eqnarray}
 D_{\frac{5}{2}} & \rightarrow & \Gamma_7+\Gamma_8, \label{2.3a} \\
 D_{\frac{7}{2}} & \rightarrow & \Gamma_6+\Gamma_7+\Gamma_8 ,
\label{2.3b} 
\end{eqnarray}
\end{mathletters}
where $\Gamma_6$, $\Gamma_7$, $\Gamma_8$ are irreducible
representations of the double cubic group $O'_h$.

Usually for CEF calculations \cite{Lea} the lowest $J$ multiplet is used
and the CEF 4$f$ Hamiltonian is expressed in terms
of Stevens equivalent operators $J_x$, $J_y$ and $J_z$.\cite{Ste,Lea}
A thorough discussion on microscopic origin of the
crystal field effects is given by Newman. \cite{New1,New2}
Here we will follow the method of Refs.~\onlinecite{Nik1,Nik2}
which allows us to calculate CEF using results of electron band structure
calculations as a starting point.
Our approach is not restricted to the $J$ multiplet with lowest
energy.
Since we may neglect conduction electrons (see Sect. I),
the problem of the CEF of one 4$f$ hole then becomes similar
to the CEF of one 4$f$ electron considered in Ref.~\onlinecite{Nik1}
with the only difference that in the following we take into
account the radial dependence of the 4$f$ hole in TmTe.
Our goal is then to diagonalize the potential $V_0^f$ and to
determine the associated 14 eigenvalues $\varepsilon_i$.

In order to proceed, we first need to know the charge distribution 
in the TmTe unit cell. 
The previous electron band structure calculation \cite{Nor} of TmTe 
in the local density approximation (LDA) does not
give such information.
We then have performed our linear augmented plane wave (LAPW)
calculations using LDA and the muffin-tin (MT) approximation. \cite{lapw}
The MT radii 2.9 and 3.1036 a.u. were chosen for Tm and Te, 
respectively.
The MT potential and density of Tm and Te have been
obtained self-consistently using a LAPW basis of $\sim170$
plane wave functions on a 20-point mesh of the irreducible
domain of the fcc Brillouin zone.
During the calculations the $f$ electrons of Tm were
treated as core states with an occupation number of 13
but were allowed to adjust
self-consistently to the conduction electron density.
We did tabulate the radial dependence ${\cal R}_f$
of the $J=7/2$ electronic 4$f$ states of Tm on a set
of 70 points with $0.111 \leq r \leq 2.971$ a.u.
Notice that ${\cal R}_f$ is obtained as an output of the
electronic band structure calculation of TmTe and thereby
deviates from the 4$f$ electronic density of a Tm atom.
The results of the charge density calculations are quoted in Table I.
%
%
%   TABLE 1
%   -------
\begin{table} 
\caption{
Angular-momentum-decomposed electronic charges $Q_l^A$ and
total charges $Q^A$
inside Tm and Te MT spheres and 
in the interstitial
region (LAPW calculations, see Ref.~31 for 
details and definitions).
\label{tab1}     } 
 
 \begin{tabular}{c | r c c c} 
  $A$ & Tm & Te & interstices \\
\tableline
$Q_s^A$    &  0.161$e$ & 1.827$e$ &  $-$  \\
$Q_p^A$    &  0.140$e$ & 4.084$e$ &  $-$ \\
$Q_d^A$    &  0.275$e$ & 0.013$e$ &  $-$ \\
\tableline
$Q^A$ & +1.509$|e|$ & +0.079$|e|$ & -1.588$|e|$ 
 \end{tabular} 
\end{table} 
Using the MT potential, we have calculated the spin-orbit
splitting. The $D_{5/2}$ states of the 4$f$ hole of Tm were found
to be separated from the $D_{7/2}$ states by $\triangle'_{so}=1.21$ eV,
which is close to the experimental value of Ref.~\onlinecite{Sur}.
In the following we consider expression (\ref{2.1}) of $V_{so}$
with $\zeta=-4112$~K.

In constructing the crystal field operator $V_{CF}$, we 
restrict ourselves to the effect of the six nearest Te neighbors in
octahedral position around a Tm$^{++}$ ion at site $\vec{n}$.
(Later we will discuss the limitation of the present approach.)
Since the crystal field has cubic symmetry, we write it in form
of a multipole expansion, following the method of Ref.~\onlinecite{Nik1},
Appendix A:
\begin{eqnarray}
 V_{CF}(\vec{n})= B_4^f\,
  \rho_{\Lambda_1}^F(\vec{n}) + B_6^f\,
  \rho_{\Lambda_2}^F(\vec{n}) . 
  \label{2.4}  
\end{eqnarray}
Here $\rho_{\Lambda_1}^F$, $\rho_{\Lambda_2}^F$ are 
the hole charge density operators,
$\Lambda_1 \equiv (l=4,A_{1g})$ refers to the lowest (non-trivial)
multipole ($l=4$) of symmetry $A_{1g}$ (unit representation of $O_h$),
while $\Lambda_2 \equiv (l=6,A_{1g})$ corresponds to the next
with $l=6$.
Explicitly, we have ($p=1,2$):
\begin{eqnarray}
 \rho_{\Lambda_p}^F(\vec{n})=\sum_{ij} c_{\Lambda_p}^f(ij)
  |i \rangle_{\vec{n}} \langle j|_{\vec{n}} ,
\label{2.5} 
\end{eqnarray}
where the basis states $|i \rangle$ are 4$f$ states, and where
\begin{mathletters}
\begin{eqnarray}
  c_{\Lambda_1}^F(ij)= \langle i | S_{l=4}^{A_{1g}} | j \rangle,
\label{2.6a} \\
  c_{\Lambda_2}^F(ij)= \langle i | S_{l=6}^{A_{1g}} | j \rangle,
\label{2.6b}
\end{eqnarray}
\end{mathletters}
$S_{\Lambda_1} \equiv S_4^{A_{1g}}$ and $S_{\Lambda_2} \equiv S_6^{A_{1g}}$
being the site symmetry adapted 
functions \cite{Bra} (SAF).
The function $S_4^{A_{1g}}$ is given explicitly by Eq.~(A.8) of 
Ref.~\onlinecite{Nik1}, the function $S_6^{A_{1g}}$ is
\begin{eqnarray}
  S_6^{A_{1g}}(\Theta,\phi)=
  \sqrt{\frac{1}{8}}\, Y_6^0(\Theta,\phi)
  -\sqrt{\frac{7}{8}}\, Y_6^{4,c}(\Theta,\phi) .
\label{2.6c} 
\end{eqnarray}
The superscript $F$ stands for quantum transitions between
4$f$ states of the hole.
Taking as basis states the functions of the irreducible representations 
$T_{1u}$, $T_{2u}$ and $A_{2u}$ of $O_h$, we obtain 
the diagonal coefficients $c^F_{\Lambda_1}(A_{2u})=-0.23505$, 
$c^F_{\Lambda_1}(T_{2u})=-0.03917$,
$c^F_{\Lambda_1}(T_{1u})=0.11752$ for $\Lambda_1$
and $c^F_{\Lambda_2}(A_{2u})=0.20118$, 
$c^F_{\Lambda_2}(T_{2u})=-0.15088$,
$c^F_{\Lambda_2}(T_{1u})=0.08382$ for $\Lambda_2$.
Notice that there is no term with $l=8$ in the expansion series
(\ref{2.4}). \cite{Lea} 
Although in principle one can consider the function \cite{Bra} 
$S_{l=8}^{A_{1g}}$,
the corresponding matrix elements $\langle i | S_{l=8}^{A_{1g}} | j \rangle$
vanish for $4f$ orbitals. 
The coefficients $B_4^f$ and $B_6^f$ are given by
\begin{mathletters}
\begin{eqnarray}
  B_4^f = 
  {\frac {6}{\sqrt{4 \pi}}}\, Q_{eff}\,e_h v_{\Lambda_1}^{F}{}_0^{\bullet} , 
  \label{2.7a} \\ 
  B_6^f = 
  {\frac {6}{\sqrt{4 \pi}}}\, Q_{eff}\,e_h v_{\Lambda_2}^{F}{}_0^{\bullet} . 
  \label{2.7b}  
\end{eqnarray}
\end{mathletters}
Here $e_h$ refers to the charge of the hole, $Q_{eff}$ to the effective
charge of Te and $v_{\Lambda_p}^{F}{}_0^{\bullet}$ is given by the
radial average of the $\Lambda_p$-th multipole of the Coulomb
potential:
\begin{mathletters}
\begin{eqnarray}
  v_{\Lambda_p}^{F}{}_{0}^{\bullet} = \int dr\,r^2\, {\cal R}_f^2(r) \,
  v_{\Lambda_p\, 0}(\vec{n},\vec{n}'; r,r') ,
\label{2.8a}
\end{eqnarray}
where ${\cal R}_f$ is the radial function \cite{Nik2} and 
\begin{eqnarray}
  & & v_{\Lambda_p\, 0}(\vec{n},\vec{n}'; r,r') = \frac{1}{\sqrt{4\pi}}
   \int d\Omega(\vec{n})\,d\Omega(\vec{n}') 
   \frac{S_{\Lambda_p}(\Omega(\vec{n})}
   {|\vec{R}(\vec{n})-\vec{R}'(\vec{n}')|} . \nonumber \\
 & & \label{2.8b}
\end{eqnarray}
\end{mathletters}
In the last integrals we have 
$\vec{R}(\vec{n})=\vec{X}(\vec{n})+\vec{r}(\vec{n})$ where $\vec{X}(\vec{n})$
is the lattice site position and $\vec{r}(\vec{n})$ the hole (electron)
radius vector, $\vec{r}=(r,\vec{\Omega})$, $\vec{\Omega}=(\Theta,\phi)$.

We take $Q_{eff}=Q^{Te}$, Table I.
In Eqs. (\ref{2.8a},b) $\vec{n}$ refers to 
a Tm lattice site, while $\vec{n}'$ - to any of its
six Te neighbors. 
Diagonalization of $V_{CF}(\vec{n})$ leads to the crystal field
term scheme without spin-orbit coupling:
\begin{eqnarray}
   \varepsilon(\Gamma)=B_4^f\, c^F_{\Lambda_1}(\Gamma)+
   B_6^f\, c^F_{\Lambda_2}(\Gamma),
\label{2.8c} 
\end{eqnarray}
where $\Gamma=A_{2u}$, $T_{2u}$ and $T_{1u}$. Our calculations yield
$B_4^f$=35.73 K, $B_6^f$=1.631 K and 
$\varepsilon(A_{2u})=-8.07$~K, $\varepsilon(T_{2u})=-1.646$~K, 
$\varepsilon(T_{1u})=4.336$~K.
Including next the spin-orbit coupling and proceeding as in appendix A
of Ref.~\onlinecite{Nik1}, we obtain the CEF term spectrum $\varepsilon_i$
of the 4$f$ hole, quoted in Table~II.
%
%
%   TABLE 2
%   -------
\begin{table} 
\caption{
Calculated CEF spectrum of a 4$f$ hole in the disordered
phase ($T>T_Q$), $\triangle \varepsilon$=14389.4 K.
\label{tab2}     } 
 
 \begin{tabular}{c c c} 
%\hline\noalign{\smallskip} 
 $i$ &  & $\varepsilon_i$ \\
\tableline
%\noalign{\smallskip}\hline\noalign{\smallskip} 
$1,2$  & $\Gamma_{7},1$ & 0 K \\
$3-6$  & $\Gamma_8,1$ & 5.81 K \\
$7,8$  & $\Gamma_6$ & 9.65 K  \\
$9,10$ & $\Gamma_7,2$ & $\triangle \varepsilon$ \\
$11-14$ & $\Gamma_8,2$ & $\triangle \varepsilon$+6.60 K 

%\noalign{\smallskip}\hline 
 \end{tabular} 
\end{table} 

If one goes beyond the present nearest neighbor approximation, 
the calculation becomes much more complex.
Considering the charge contributions from the twelve neighboring 
Tm sites and $Q^{int}=-1.588|e|$ from the interstitial region, we find
that the crystal field splittings between energy levels increase by roughly 
an order of magnitude in comparison with the values of Table II,
$\varepsilon(\Gamma_8,1)-\varepsilon(\Gamma_7,1)$=62.9 K,
$\varepsilon(\Gamma_6)-\varepsilon(\Gamma_7,1)$=102.5 K.
The increase is mainly caused by a larger positive value of $Q_{eff}$
in Eqs.~(\ref{2.7a},b) when
the homogeneous electron charge distribution
in the interstitial region is taken into account. \cite{full}
Notice, however, that the sequence $\Gamma_7-\Gamma_8-\Gamma_6$ with 
$\Gamma_7$ as a ground state is conserved. 
The other remote shells of neighbors have been found to produce little
changes on the calculation since the corresponding integrals
vary like \cite{Hei} $|\vec{R}(\vec{n})-\vec{R}(\vec{n}')|^{-5}$ for 
$S^{A_{1g}}_{\Lambda_1}$ and 
$|\vec{R}(\vec{n})-\vec{R}(\vec{n}')|^{-7}$ for 
$S^{A_{1g}}_{\Lambda_2}$.
The magnitude of the CEF splittings in the latter approach is
reduced by several effects: screening due to polarization of
conduction electrons, \cite{Nik2} which are still present on each Tm site
(see Table I), and an inhomogeneous charge distribution in the
interstitial volume. 
The first effect is due to 
the coupling of 4$f$ localized electrons with 5$d$ conduction electron 
states which was discussed by Newman. \cite{New2} 
In our model the interactions with
the conduction electrons are given by Eq.~(4.10) of Ref. \onlinecite{Nik2}.
However, such calculation would require a self-consistent procedure \cite{Nik2}
for 4$f$ and conduction electrons included in an electron band structure
scheme and is beyond the scope 
of the present work which is focused on quadrupolar ordering. 
Some authors \cite{Bar,New1} have emphasized the role of
covalent mixing which involves one-electron matrix elements
between the 4$f$ electrons with 5$d$ electrons on other atoms.
Notice that such effects imply nonorthogonal basis
functions on neighboring sites. In LAPW \cite{Sin} and 
linear muffin-tin orbital (LMTO) \cite{And} 
methods of electron band structure calculations there is no overlap
between MT basis functions belonging to different sites,
while 4$f$ electronic wave functions are confined inside MT spheres.
The covalent effect which is associated with anisotropic character
of chemical bonding then may be described by inhomogeneity of electron
charge distribution in the interstitial region and inside MT spheres.
Since there is a large portion of electronic charge situated
in the interstices ($Q^{int}$) it is likely that the effect of the
charge inhomogeneity is appreciable.

In summary we here obtain the sequence $\Gamma_7-\Gamma_8-\Gamma_6$ with 
$\Gamma_7$ as ground state, in agreement with results from
M\"{o}ssbauer spectroscopy \cite{Tri} and ultrasonic
velocity measurements. \cite{Kas}
% but in disagreement with other experimental studies. \cite{Ott,Cle}
Although the present calculation of CEF in incomplete
our estimations show that the sequence is likely to be conserved
in a full treatment of the problem.
%Here we would like to remark that if a negative charge of Te is taken
%(complete charge transfer), then the order of CEF is reversed, $i.e.$
%the sequence becomes $\Gamma_6-\Gamma_8-\Gamma_7$ with ground state
%$\Gamma_6$. 
%Even in this case  the $\Gamma_8$ quartet is not realized as
%ground state.
The experimental identification of the ground state as $\Gamma_8$
is in contradiction with the present calculations.
The discrepancy can arise due to strong fluctuations of quadrupole
density which occurs at temperature $T \sim 2-3$~K and affect 
the experimental results. 
Indeed, the presence of quadrupolar fluctuations have been
found at 3~K in M\"{o}ssbauer studies of TmTe. \cite{Tri}
It is necessary to analyze both the experimental conditions
and methods of obtaining the CEF levels and of identifying the states.
Further investigations and full calculations of CEF are needed 
in order to clarify the issue.
On the other hand, we want to stress that theoretically
a quadrupole order can occur even
if the ground state of the 4$f$ hole 
is $\Gamma_7$ (or $\Gamma_6$) \cite{Nik1,Kas}
while $\Gamma_8$ remains an excited state. 
%It is worth to note
%that a quadrupolar ordering has been found experimentally in
%ytterbium monopnictides where the ground state
%at $T>T_Q$ is $\Gamma_6$. \cite{Kel}

%------------- III --------------------------> 
\section {Phase Transition} 
\label{sec:CL} 

Here we will discuss the antiferroquadrupolar phase transition 
in TmTe.
In fact this phase transition is a structural one,
with the concomitant symmetry change cubic$\rightarrow$monoclinic.
In the following we will continue to exploit the duality electron-hole
between Ce and TmTe.
In Ref.~\onlinecite{Nik1} we have shown that the Coulomb interaction operator
between 4$f$ electrons (holes) on a fcc lattice is obtained as
a double multipole expansion 
\begin{eqnarray}
  U^{ff}={\frac{1}{2}} {\sum_{\vec{n} \vec{n}'}}' \sum_{\Lambda \Lambda'}
  \rho_{\Lambda}^F(\vec{n}) \,
  v_{\Lambda \Lambda'}(\vec{n}-\vec{n}') \,
  \rho_{\Lambda'}^F(\vec{n}') .
\label{3.1} 
\end{eqnarray}
Here the expansion coefficients $v_{\Lambda \Lambda'}$ are given by
\begin{mathletters}
\begin{eqnarray}
  v_{\Lambda \Lambda'}(\vec{n}-\vec{n}') &=&
  \int \! dr\, r^2 \int \! dr'\, {r'}^2\,  \nonumber \\
  & &\times {\cal R}_f^2(r) {\cal R}_f^2(r')\,
  v_{\Lambda \Lambda'}(\vec{n},\vec{n}';\, r,r') ,
\label{3.2a} 
\end{eqnarray}
where
\begin{eqnarray}
 & & v_{\Lambda \Lambda'}(\vec{n}-\vec{n}';r,r') =
 \int d\Omega(\vec{n})\, d\Omega(\vec{n}')
 \frac{S_{\Lambda}(\hat{n})\,S_{\Lambda}(\hat{n}')}
 {|\vec{R}(\vec{n})-\vec{R}'(\vec{n}')|} . \nonumber \\
 & & \label{3.2b} 
\end{eqnarray}
\end{mathletters}
For details on the radial average in Eq.~(\ref{3.2a}), see 
Ref.~\onlinecite{Nik2}.
Here $S_{\Lambda}(\hat{n})$, $\hat{n} \equiv (\Theta(\vec{n}),\phi(\vec{n}))$,
are site symmetry adapted functions \cite{Bra},
$\Lambda$ stands for $(l,\tau)$, where $l$ accounts for the angular
dependence of the multipolar expansion and $\tau=(\Gamma,k)$,
$\Gamma$ denoting the irreducible representation of the site
point group and $k$ labeling the rows of $\Gamma$.
In our case, $l=2$ (quadrupoles) and $\Gamma$ stands for the
representations $T_{2g}$ ($k=1-3$) and $E_g$ ($k=1,2$) of the cubic
site group $O_h$. 
The corresponding SAFs $S_{T_{2g}}^k$ and $S_{E_g}^k$ are given
by Eqs.~(2.16) and (2.15) of Ref.~\onlinecite{Nik1}.
The quantity $\rho_{\Lambda}^F(\vec{n})$ stands for
the multipolar density
\begin{eqnarray}
 \rho_{\Lambda}^F(\vec{n})=\sum_{ij} c^F_{\Lambda}(ij)
 |i\rangle _{\vec{n}}\langle j|_{\vec{n}} ,
\label{3.2} 
\end{eqnarray}
with
\begin{eqnarray}
  c_{\Lambda}^F(ij) =  
  \int \! d\Omega \, \langle i|\hat{n}\rangle  
  S_{\Lambda}(\hat{n}) \langle \hat{n}|j\rangle .
\label{3.3} 
\end{eqnarray}
 
Introducing Fourier transforms
\begin{mathletters}
\begin{eqnarray}
 & &\rho_{\Lambda}^F(\vec{q})= {\frac {1}{\sqrt{N}}}
 \sum_{\vec{n}} e^{i\vec{q} \cdot \vec{X}(\vec{n})}
 \rho_{\Lambda}^F(\vec{n}),  \label{3.4a} \\  
 & &v_{\Lambda \Lambda'}(\vec{q})=  
 {\sum_{\vec{h} \neq 0}}' 
  e^{i\vec{q} \cdot \vec{X}(\vec{h})} 
  v_{\Lambda \Lambda'}(\vec{h}), 
     \label{3.4b}  
\end{eqnarray}
\end{mathletters}
where $\vec{q}$ is the wave vector, we get for the quadrupole-quadrupole
interaction
\begin{eqnarray}
 U^{ff}_{QQ}={\frac {1}{2}} \sum_{\vec{q}} \sum_{\Lambda \Lambda'}
 \rho_{\Lambda}^F(\vec{q})^{\dagger}  
 v_{\Lambda \Lambda'}(\vec{q}) \, 
 \rho_{\Lambda'}^F(\vec{q}).
\label{3.5} 
\end{eqnarray}

The $5 \times 5$ matrix $v_{\Lambda \Lambda'}$
is given by the expressions (A1, A6) and (A7)
of Ref.~\onlinecite{Mic1}. This matrix has negative eigenvalues
at some points of the Brillouin zone (BZ).
The largest negative eigenvalues have been found at the $X$ and $L$
points of the BZ.
Since the superstructure reflections 
have been
found by neutron-diffraction experiments \cite{Lin2} on TmTe 
at the $L$ point of the BZ, $\vec{q}_L=(2\pi/a)(1/2,1/2,1/2)$,
we limit in the following our considerations to the $L$ point.
%It is possible that condensation of the 
%quadrupole mode $X_2^+$ at the $X$ point of BZ is realized
%in the tetragonal structure of TmTe which is observed
%under pressure 5.7 GPa \cite{Mat1,Hea,Tan}.

There are four arms of the star $^*\vec{q}_L$ which we label by
$\vec{q}_L^{\,1}=(1/2,1/2,1/2)$, $\vec{q}_L^{\,2}=(-1/2,1/2,1/2)$,
$\vec{q}_L^{\,3}= (1/2,-1/2,1/2)$ and 
$\vec{q}_L^{\,4}= (-1/2,-1/2,1/2)$, in units $(2\pi/a)$.
At $\vec{q}=\vec{q}_L^{\;i}$, $i=1-4$,
the eigenvalue spectrum of the quadrupole
matrix $v_{\Lambda \Lambda'}$ is the same and for
simplicity we consider the arm $\vec{q}_L^{\,1}$. 
Notice that at the $L$ point there is a coupling between components
of $T_{2g}$ and $E_g$ symmetry 
(see (A7) of Ref.~\onlinecite{Mic1}). We write
\begin{eqnarray}
& &v(\vec{q}_L^{\,1})=
 \left[ \begin{array}{c c}
 v^{TT} & v^{TE}  \\ 
 v^{ET} & \hat{0}   
\end{array} \right] ,  
 \label{3.6} 
\end{eqnarray}
where $\hat{0}$ stands for the $2 \times 2$ zero matrix,
$v^{TT}$ describes the $3 \times 3$ matrix between components
of $T_{2g}$ symmetry,
\begin{eqnarray}
& &v^{TT}(\vec{q}_L^{\,1})=
-4 \left[ \begin{array}{c c c}
 0 & \beta & \beta \\ 
 \beta & 0 & \beta \\
 \beta & \beta & 0 
\end{array} \right] ,  
 \label{3.7} 
\end{eqnarray}
and $v^{TE}$ stands for the $3 \times 2$ $T_{2g}-E_g$ coupling matrix,
\begin{eqnarray}
& &v^{TE}(\vec{q}_L^{\,1})=
-4 \left[ \begin{array}{c r}
 \lambda & \nu  \\ 
 \lambda & -\nu  \\
 \mu & 0  
\end{array} \right] ,  
 \label{3.8} 
\end{eqnarray}
and $v^{ET}=(v^{TE})^{\dagger}$.
The elements $\beta$, $\lambda$, $\mu$ and $\nu$ are obtained
by integrals of the type (\ref{3.2a},b).

Diagonalizing the matrix $v^{TT}$ we obtain the
eigenvalues $-8\beta$, 4$\beta$, 4$\beta$ and eigenvectors
\begin{mathletters}
\begin{eqnarray}
 & &{S'}_{T_{2g}}^1(\vec{q}_L^{\,1})=\frac{1}{\sqrt{3}}
 (S_{T_{2g}}^1(\vec{q}_L^{\,1})+S_{T_{2g}}^2(\vec{q}_L^{\,1})+
 S_{T_{2g}}^3(\vec{q}_L^{\,1})), \\
 & &{S'}_{T_{2g}}^2(\vec{q}_L^{\,1})=\frac{1}{\sqrt{2}}
 (S_{T_{2g}}^1(\vec{q}_L^{\,1})-S_{T_{2g}}^2(\vec{q}_L^{\,1})), \\
 & &{S'}_{T_{2g}}^3(\vec{q}_L^{\,1})=\frac{1}{\sqrt{6}}
 (S_{T_{2g}}^1(\vec{q}_L^{\,1})+S_{T_{2g}}^2(\vec{q}_L^{\,1})
 -2 S_{T_{2g}}^3(\vec{q}_L^{\,1})). 
 \label{3.9} 
\end{eqnarray}
\end{mathletters}
In the basis ${S'}_{T_{2g}}^1(\vec{q}_L^{\,1})$, 
${S'}_{T_{2g}}^2(\vec{q}_L^{\,1})$, 
$S_{E_{g}}^2(\vec{q}_L^{\,1})$, ${S'}_{T_{2g}}^3(\vec{q}_L^{\,1})$, 
$S_{E_{g}}^1(\vec{q}_L^{\,1})$ 
the matrix $v(\vec{q}_L^{\,1})$ becomes block-diagonal,
\begin{eqnarray}
& &v(\vec{q}_L^{\,1})= -4 
 \left[ \begin{array}{c c c c c}
  2\beta &  0     &  0          & 0 & 0 \\ 
  0      & -\beta & \sqrt{2}\nu & 0 & 0 \\
  0 & \sqrt{2}\nu & 0           & 0 & 0 \\
  0 & 0 & 0 & -\beta & -\sqrt{\frac{3}{2}}\mu \\
  0 & 0 & 0 & -\sqrt{\frac{3}{2}}\mu &  0
\end{array} \right] .  
 \label{3.10} 
\end{eqnarray}
We find its eigenvalues, of which 
$\lambda_L^1=-2(-\beta+\sqrt{\beta^2+8\nu^2})$
and $\lambda_L^2=2(\beta+\sqrt{\beta^2+8\nu^2})$ are double degenerate
while $\lambda_L^3=-8\beta$ is non degenerate.
From numerical calculations we obtain
$\beta=-33.54$~K, $\mu=-29.05$~K, $\lambda=14.53$~K,
$\nu=-25.16$~K and $\lambda_L^1=-224.4$~K,
$\lambda_L^2=90.3$~K and $\lambda_L^3=268.3$~K.
These results are in agreement with the symmetry relations
$2\nu=\sqrt{3}\mu$, $2\lambda=-\mu$, that hold for quadrupole-quadrupole
Coulomb interactions on a fcc lattice.
The lowest eigenvalue $\lambda_L^1$ has the eigenvectors
\begin{mathletters}
\begin{eqnarray}
 S^{(1)}(\vec{q}_L^{\,1}) &=& 
 -0.5972 (S_{T_{2g}}^1  - S_{T_{2g}}^2 ) +
 0.5356\, S_{E_{g}}^2 , \label{3.11a} \\
 S^{(2)}(\vec{q}_L^{\,1})&=& +0.3448
 (S_{T_{2g}}^1 +S_{T_{2g}}^2 )
 -0.6895\, S_{T_{2g}}^3  \nonumber \\
 & &+ 0.5356\, S_{E_{g}}^1  , 
 \label{3.11b} 
\end{eqnarray}
\end{mathletters}
where we omit the arguments $\vec{q}_L^{\,1}$ on the right hand sides.
In addition, we consider the corresponding functions in real
space,
\begin{mathletters}
\begin{eqnarray}
 S^{(1)}(\Omega)&=& 
 -0.5972 (S_{T_{2g}}^1(\Omega)  - S_{T_{2g}}^2(\Omega) )  \nonumber \\
 & &+
 0.5356\, S_{E_{g}}^2(\Omega) , \label{3.12a} \\
 S^{(2)}(\Omega)&=& +0.3448
 (S_{T_{2g}}^1(\Omega) +S_{T_{2g}}^2(\Omega) ) \nonumber \\
 & &-0.6895\, S_{T_{2g}}^3(\Omega) + 0.5356\, S_{E_{g}}^1(\Omega)  . 
 \label{3.12b} 
\end{eqnarray}
\end{mathletters}
These two functions are shown in Fig.~1a,b.
We investigate their transformational properties in detail
in Appendix A.
%------------------------------------------------------
%    FIGURE 1
%------------------------------------------------------
\begin{figure} 
%\vspace{-1.1cm}
\centerline{
\epsfig{file=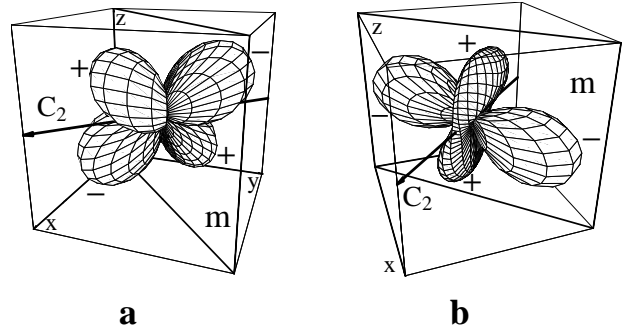,width=0.52\textwidth}
} 
\vspace{-7cm}
% -----------> Figure Caption
\caption{
 Schematic pictures of two variants of quadrupole order parameter
 (quadrupole density);
 $m$ is the mirror plane, $C_2$ is the rotation axis.
 a) $S^{(1)}$, Eq.~(\ref{3.12a}), which
 leads to $C2/c$, Eq.~(\ref{3.16a});
 b) $S^{(2)}$, Eq.~(\ref{3.12b}), which
 leads to $C2/m$, Eq.~(\ref{3.16b}).
} 
\label{fig1} 
\end{figure} 

The quadrupolar densities which correspond to the functions 
$S^{(\alpha)}$, $\alpha=1,2$, are given by the expression
\begin{mathletters}
\begin{eqnarray}
   \rho_{\alpha}^F(\vec{n})=\sum_{ij} |i \rangle_{\vec{n}}\, c_{\alpha}^F(ij)
   \langle j |_{\vec{n}}
 \label{3.19a}
\end{eqnarray}
with
\begin{eqnarray}
 c_{\alpha}^F(ij)=\langle i | S^{(\alpha)} | j \rangle =
 \int d\Omega \langle i| \hat{n} \rangle S^{(\alpha)}(\hat{n})
 \langle \hat{n} | j \rangle .
 \label{3.19b}
\end{eqnarray}
\end{mathletters}
(Compare with expressions (\ref{3.2}), (\ref{3.3}).)

The functions $S^{(\alpha)}$ belong
to the two dimensional small representation $E_g$
of the little group ${\bar 3}m$ ($D_{3d}$) of 
$^*\vec{q}_L$ ($\hat{\tau}^5$ representation in
Kovalev's notation \cite{Kov}).
The irreducible representation of the
space group $Fm{\bar 3}m$ comprises eight such functions,
with two functions from four arms of $^*\vec{q}_L$, that is, 
$S^{(1)}(\vec{q}_L^{\,1})$, $S^{(2)}(\vec{q}_L^{\,1})$;
$S^{(1)}(\vec{q}_L^{\,2})$, $S^{(2)}(\vec{q}_L^{\,2})$;
$S^{(1)}(\vec{q}_L^{\,3})$, $S^{(2)}(\vec{q}_L^{\,3})$
and $S^{(1)}(\vec{q}_L^{\,4})$, $S^{(2)}(\vec{q}_L^{\,4})$.
In principle, there are many possibilities for
condensations schemes at $^*\vec{q}_L$ involving one, two, three
or four arms. \cite{Lau,Sto}
Experimentally, reflections associated with all four components
of the star $^*\vec{q}_L$ were clearly observed \cite{Lin2}
and had different intensities even in small applied magnetic fields.
On this basis it was concluded in Ref.~\onlinecite{Lin2} that each
arm of $^*\vec{q}_L$ is associated with a domain.
We then limit our consideration to the case where a single
arm, say $\vec{q}_L^{\,1}$, is involved in the symmetry lowering
which occurs due to the quadrupolar ordering.
In such case the following two condensation schemes are possible \cite{Sto}:
\begin{mathletters}
\begin{eqnarray}
 & &Fm{\bar 3}m: \;\;L_3^+\; 
 [\langle \rho_{1}^F(\vec{q}_L^{\,1}) \rangle = \sqrt{N} \rho_1]
 \rightarrow C2/c\,(Z=2)  ,  \nonumber \\
 & & \label{3.16a} \\
 & &Fm{\bar 3}m: \;\;L_3^+\; 
 [\langle \rho_{2}^F(\vec{q}_L^{\,1}) \rangle = \sqrt{N} \rho_2]
 \rightarrow C2/m\,(Z=2) .  \nonumber \\
 & & \label{3.16b}
\end{eqnarray}
\end{mathletters}
Here $\rho_{\alpha}^F(\vec{q})$ stands for the Fourier transform of
$\rho_{\alpha}^F(\vec{n})$, $\langle ... \rangle$ denotes a thermal
average and $\rho_{\alpha}$ are the order parameter amplitudes.
Correspondingly, in real space we obtain
\begin{eqnarray}
  \langle \rho_{\alpha}^F(\vec{n}) \rangle=\rho_{\alpha} 
  \cos(\vec{q}_L^{\,1} \cdot \vec{X}(\vec{n}))  , \;\;\;\; \alpha=1,2.
 \label{3.17}
\end{eqnarray}
Both structures are monoclinic, with the mirror plane $[1 {\bar 1} 0]$,
see also Figs. 1a,b.
As follows from Eq. (\ref{3.17}) $\langle \rho^F_1(\vec{n}) \rangle$
and $\langle \rho^F_2(\vec{n}) \rangle$ change from $+ \rho_1$
to $- \rho_1$ and from $+ \rho_2$ to $- \rho_2$, respectively,
along the $[110]$ direction. The resulting pattern as well as the monoclinic
unit cell are shown in Fig. 2.
%------------------------------------------------------
%    FIGURE 2
%------------------------------------------------------
\begin{figure} 
\vspace{-1.5cm}
\resizebox{0.44\textwidth}{!}
{ 
\hspace{-1.3cm}\includegraphics{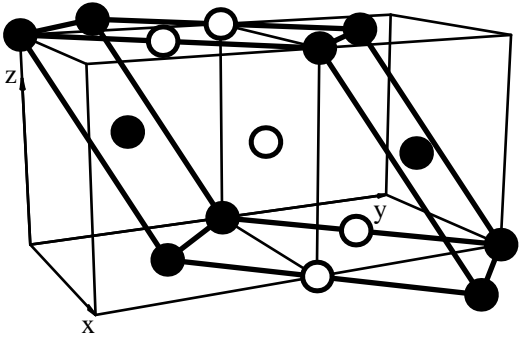} 
} 
\vspace{-4.5cm}
% -----------> Figure Caption
\caption{
 Monoclinic unit cell in respect to cubic system of axes.
 Black and white circles refer to Tm sites where
 in case of $C2/c$ structure $\langle \rho^F_1 \rangle$
 (in case of $C2/m$ $\langle \rho^F_2 \rangle$) is taken
 with the sign $+$ for black and $-$ for white circles. 
} 
\label{fig2} 
\end{figure} 
We speak of an antiferro-quadrupolar order.
At $\vec{q}_L^{\,1}$ there are still three variants of condensations
of the type (\ref{3.16a}) and three of the type (\ref{3.16b}).
For example, we consider the condensation in $C2/c$, Eq.~(\ref{3.16a}).
The three variants involve condensations of quadrupolar functions
${S'}^{(1)}$ which are obtained from $S^{(1)}$ through rotations
by the angle 0, $2\pi/3$ and $4\pi/3$ about the cubic axis $[111]$.
Notice that this functions can be expressed in terms of a linear
combinations of $S^{(1)}$, $S^{(2)}$ since they form a basis
of the little group ${\bar 3}m$ ($D_{3d}$) of 
$\vec{q}_L^{\, 1}$. Otherwise, the consideration is the same as for
$S^{(1)}$, $S^{(2)}$. The corresponding monoclinic unit cells
are obtained from that in Fig.~2 through the same rotations by
the angle 0, $2\pi/3$ and $4\pi/3$. For example, for the
rotation by $2\pi/3$ the basal $(xy)$ plane (see Fig. 2)
transforms to $(yz)$ while the monoclinic mirror plane $[1 {\bar 1} 0]$
becomes $[0 1 {\bar 1}]$.
These three variants correspond to the so-called ``S domains"
which have been observed in neutron diffraction experiments \cite{Lin2}
for a given arm of $^*\vec{q}_L$.
However, the total number of domains for the condensation to $C2/c$
is twelve.
The same holds for the second condensation scheme, Eq.~(\ref{3.16b}),
to the $C2/m$ monoclinic structure.
We conclude that on the basis of data from neutron diffraction \cite{Lin2}
on TmTe
it is not possible to deduce which of the two ordered structures 
actually occurs in TmTe. 
Both structures are monoclinic and lead to the domain
structure observed in experiment. \cite{Lin2}
On the other hand, from our theoretical analysis of coupling matrices
we cannot rule out one structure in favor of the other.
In the next section we show that the two types of quadrupolar ordering can be
distinguished by lattice displacements which accompany the
transition.
The condensation scheme (\ref{3.16a}) corresponds to 
that given by Eq. (3.10a) of
Ref.~\onlinecite{Par}, where two complex basis functions of the $\hat{\tau}^5$
irreducible representation are used.

In the following we study the thermodynamics of the quadrupolar
phase transitions.
Taking into account the first condensation
scheme, Eq. (\ref{3.16a}), we obtain for the quadrupole coupling at
a site $\vec{n}_1$ for one sublattice ($\{ \vec{n}_1 \}$)
\begin{eqnarray}
   U^{ff}_{QQ}(\vec{n}_1)= \lambda_L^{1} \rho_1 \rho_1^F(\vec{n}_1),
 \label{3.18}
\end{eqnarray}
where $\rho_1$ is the order parameter amplitude and where
$\rho_1^F(\vec{n}_1)$ is the quadrupolar density operator.
The mean field Hamiltonian reads
\begin{eqnarray}
  H^{MF}(\vec{n}_1)=U^{ff}_{QQ}(\vec{n}_1)+V_0^f(\vec{n}_1) ,
 \label{3.21}
\end{eqnarray}
where $V_0^f$, Eq.~(\ref{2.2}), describes the crystal field and the spin-orbit
coupling.
Starting with $H^{MF}$ and
using methods which have been developed for molecular crystals \cite{Mic}, 
we obtain the following
approximate expression for the Landau free energy:
\begin{eqnarray}
   {\cal F}/N={\cal F}_0/N+  
    A \rho_1^2 + B \rho_1^4 .
 \label{3.22}
\end{eqnarray}
Here ${\cal F}_0$ is the free energy in the disordered phase
\begin{eqnarray}
 {\cal F}_0/N = -T \ln Z_0 ,
 \label{3.24}
\end{eqnarray}
where
\begin{eqnarray}
 Z_0 = \sum_{i=1}^{14} e^{-\varepsilon_i/T}   \label{3.23c}
\end{eqnarray}
is the sum of states and $\varepsilon_i$ are the crystal field energy 
levels. 
The expansion coefficients in Eq.~(\ref{3.22}) are
\begin{mathletters}
\begin{eqnarray}
  & & A = \frac{1}{2} 
       \left( \frac{T}{x^{(2)}}+\lambda_L^1 \right) , \label{3.n23a} \\ 
  & & B = \frac{T}{24 [x^{(2)}]^2} 
      \left(  3-\frac{x^{(4)}}{[x^{(2)}]^2} \right) ,
 \label{3.n23b}
\end{eqnarray}
\end{mathletters}
where 
\begin{mathletters}
\begin{eqnarray}
  & & x^{(2)}=\sum_{ij} c^F_{1}(ij)\, c^F_{1}(ji)\, 
             e^{-\varepsilon_i/T}/Z_0 ,
 \label{3.23a} \\
  & & x^{(4)}=\sum_{ijhl} c^F_{1}(ij)\, c^F_{1}(jh)\, 
              c^F_{1}(hl)\, c^F_{1}(li)\, e^{-\varepsilon_i/T}/Z_0 .
 \label{3.23b} 
\end{eqnarray}
\end{mathletters}

The calculation of $x^{(2)}$ and $x^{(4)}$ 
requires the knowledge of the crystal field. 
Using the values of $\varepsilon_i$ from Table II and the corresponding
calculated eigenvectors,
we obtain the results quoted in Table III.
%
%
%   TABLE 3
%   -------
\begin{table} 
\caption{
 Calculated parameters of the Landau free
 energy expansion, see text for details.
\label{tab3}     } 
 
 \begin{tabular}{ c c c c } 
   $x^{(2)}$ & $x^{(4)}$ & $B/T$ & $k$ \\
\tableline
   0.0219 & 0.000768 & 121.5 & 0.188  
 \end{tabular} 
\end{table} 
Since there is no third order cubic invariant in expression (\ref{3.22})
and since $B>0$, the phase transition
is of second order, with the transition temperature given by
\begin{eqnarray}
 T_c = x^{(2)} \, | \lambda_L^1 | 
 \label{3.25}
\end{eqnarray}
and the order parameter amplitude given at $T<T_c$ by
\begin{eqnarray}
  \rho_1(T)=\pm \sqrt{-\frac{A}{2B}}=\pm \sqrt{k \frac{T_c-T}{T}} ,
 \label{3.26}
\end{eqnarray}
where
\begin{eqnarray}
 k = \frac{12 (x^{(2)})^3}{3(x^{(2)})^2-x^{(4)}} .
 \label{3.27}
\end{eqnarray}
With $\lambda_L^1=-224.4$~K we find $T_c=4.9$.  
This value is more than twice the experimental
temperature $T_Q=1.8$~K. 
We ascribe the origin of the discrepancy to the
screening effect of conduction electrons from Tm and Te sites and 
from the interstitial region. The question may arise why
in case of cerium the polarization of conduction electrons leads
to an increase of transition temperature \cite{Nik2}, while in case of TmTe
it has the opposite effect. We recall that in Ce quadrupoles 
constructed from conduction electrons
are in close contact and their ordering greatly reduces the repulsion between
conduction electrons. In TmTe 4$f$ holes and conduction electrons 
around Tm sites are at larger distances
and the polarization of conduction electrons merely reduces
the resulting effective quadrupolar value.

Finally we mention that with Eqs. (\ref{3.18})-(\ref{3.25})
we readily obtain the corresponding expressions for the
second condensation scheme (\ref{3.16b}) by replacing the index 1 in $\rho_1$
and $c_{1}(ij)$ by the index 2.
The numerical values of $x^{(2)}$, $x^{(4)}$, $B$
and $T_c$ remain the same and therefore no distinction between
$C2/c$ and $C2/m$ can be made at this point.

%------------- IV -------------------------> 
\section {Lattice distortions} 
\label{sec:CC}

The quadrupolar ordering and symmetry lowering is accompanied
by a distortion of the cubic lattice.
Such effects are known to occur in molecular crystals 
(see for a review Ref. \onlinecite{Lyn})
and our present treatment \cite{Nik1,Nik2} was inspired by
the theory of orientational order in molecular solids. \cite{Lam,Lyn}
 
We consider the Tm atoms located on a non rigid fcc lattice
and denote the lattice displacement of the Tm nucleus at site $\vec{n}$
by $\vec{u}(\vec{n})$. For the 4$f$ hole coordinates we have
\begin{eqnarray}
  {\vec R}({\vec n})={\vec X}({\vec n})+{\vec r}({\vec n})+
   {\vec u}({\vec n}) ,
\label{4.1} 
\end{eqnarray}
where ${\vec X}({\vec n})$ stands for equilibrium nuclear position.
We expand the intersite potential (\ref{3.1}) in terms of atomic
lattice displacements. The first order correction to the potential reads
\begin{eqnarray}
  U_{QQT}^{(\alpha)} &=& {\frac {1}{2}} {\sum_{\vec{n} \vec{n}'}}' \sum_{\nu}
   v'_{\nu}{}^{(\alpha)}(\vec{n}-\vec{n}';r,r') \nonumber \\
  & & \times 
 S^{(\alpha)}(\hat{n}) \, S^{(\alpha)}(\hat{n}')\,
  \left[ u_{\nu}(\vec{n}) -u_{\nu}(\vec{n}') \right] ,
\label{4.2} 
\end{eqnarray}
where
\begin{eqnarray}
 v'_{\nu}{}^{(\alpha)}(\vec{n}-\vec{n}';\;r,r') 
   &=&  \int \! d\Omega(\vec{n}) \int \! d\Omega(\vec{n}')\, 
  S^{(\alpha)}(\hat{n})\, S^{(\alpha)}(\hat{n}')  
   \nonumber \\
   & \times &  \left.
  {\frac {\partial}{\partial X_{\nu}(\vec{n})}}
  {\frac{1}{|\vec{R}(\vec{n})-\vec{R}'(\vec{n}')|}} \right|_{\vec{u}=0} .
\label{4.3} 
\end{eqnarray}
Here the index $\alpha$ ($\alpha=1,2$) corresponds to the two variants
of antiferro- quadrupole ordering, Eqs.~(\ref{3.16a},b). 
We recall that $\vec{n}-\vec{n}'$ stands for 
$\vec{X}(\vec{n}-\vec{n}') \equiv \vec{X}(\vec{\kappa})$.
We take the average over the radial dependence of the 4$f$ hole:
\begin{eqnarray}
& & v'_{\nu}{}^{(\alpha)}({\vec \kappa}) =\int dr\, r^2 \int  dr'\, {r'}^2\, 
  {\cal R}_f^2(r)\, {\cal R}_f^2(r')\,
  v'_{\nu}{}^{(\alpha)}(\vec{\kappa};\;r,r'). \nonumber \\
& & \label{4.5a} 
\end{eqnarray}
One has the symmetry relation
\begin{eqnarray}
 v'_{\nu}{}^{(\alpha)}({\vec \kappa})=
 -v'_{\nu}{}^{(\alpha)}({-\vec \kappa})
\label{4.5b} 
\end{eqnarray}
on the fcc lattice.
In the following $\vec{\kappa}$ labels the twelve nearest neighboring Tm sites 
around a Tm ion taken as origin.
Proceeding as in Ref.~\onlinecite{Nik1} we rewrite the expression (\ref{4.2})
as an operator in the space of the 4$f$ hole:
\begin{eqnarray}
  U_{QQT}^{(\alpha)} &=& {\frac {1}{2}} {\sum_{\vec{n} \vec{n}'}}' \sum_{\nu}
   v'_{\nu}{}^{(\alpha)}(\vec{n}-\vec{n}') \nonumber \\
  & & \times 
 \rho^F_{\alpha}(\vec{n}) \, \rho^F_{\alpha}(\vec{n}') \,
  \left[ u_{\nu}(\vec{n}) -u_{\nu}(\vec{n}') \right] ,
\label{4.5c} 
\end{eqnarray}
where $\rho^F_{\alpha}(\vec{n})$ is defined by expressions (\ref{3.19a},b).
Transforming to Fourier space we find
\begin{eqnarray}
 U_{QQT}^{(\alpha)} =  i \sum_{\vec k} \sum_{\vec p}  
  v'_{\nu}{}^{(\alpha)}({\vec k},{\vec p}) 
  \rho^F_{\alpha}(-{\vec p}-{\vec q}) \rho^F_{\alpha}({\vec p}) \, 
   u_{\nu}({\vec k}) ,
\label{4.4} 
\end{eqnarray}
where
\begin{eqnarray}
  v'_{\nu}{}^{(\alpha)}({\vec k},{\vec p}) = (Nm)^{-1/2}
  \sum_{\vec \kappa} v'_{\nu}{}^{(\alpha)}({\vec \kappa}) 
   \nonumber \\
   \times
   \cos [({\vec p}+{\frac {\vec k}{2}}) \cdot {\vec X}(\vec \kappa)]
   \sin[{\frac {{\vec k} \cdot {\vec X}({\vec \kappa})}{2}}].
\label{4.5} 
\end{eqnarray}
In order to obtain the free energy contribution from $U_{QQT}^{(\alpha)}$,
we take the long wavelength limit $\vec{k} \rightarrow 0$ and retain
only linear terms in $\vec{k}$.
We then consider $\vec{p}$ near $\vec{q}_L^{\,1}$ and apply the
condensation schemes (\ref{3.16a},b).
Finally we replace the displacements $u_{\nu}(\vec{k})$ by their
instantaneous thermal expectation values $\langle u_{\nu}(\vec{k}) \rangle$.
After some algebra we obtain
\begin{eqnarray}
  & & F_{QQT}^{(\alpha)} = \frac{N}{2\sqrt{Nm}} (\rho_{\alpha})^2 
  \nonumber \\ 
  & & \times   \sum_{\nu=x,y,z} 
  [ik_x \Lambda^{(\alpha)}_{x \nu}  + ik_y \Lambda^{(\alpha)}_{y \nu}
  + ik_z \Lambda^{(\alpha)}_{z \nu} ]
 \langle u_{\nu}(\vec{k}) \rangle ,
\label{4.7a} 
\end{eqnarray}
where $\rho_{\alpha}$ is the order parameter amplitude.
In (\ref{4.7a}) we have defined
\begin{mathletters}
\begin{eqnarray}
 & & \Lambda^{(\alpha)}_{x \nu} = -a [ v'_{\nu}{}^{(\alpha)}(3) +
 v'_{\nu}{}^{(\alpha)}(2) + v'_{\nu}{}^{(\alpha)}(6) +
 v'_{\nu}{}^{(\alpha)}(5) ] , \nonumber \\  
 & & \label{4.14a}  \\ 
 & & \Lambda^{(\alpha)}_{y \nu} = -a [ v'_{\nu}{}^{(\alpha)}(1) +
 v'_{\nu}{}^{(\alpha)}(3) + v'_{\nu}{}^{(\alpha)}(4) -
 v'_{\nu}{}^{(\alpha)}(6) ]  ,  \nonumber \\
 & & \label{4.14b}  \\ 
 & & \Lambda^{(\alpha)}_{z \nu} = -a [ v'_{\nu}{}^{(\alpha)}(1) +
 v'_{\nu}{}^{(\alpha)}(2) - v'_{\nu}{}^{(\alpha)}(4) -
 v'_{\nu}{}^{(\alpha)}(5) ] .   \nonumber \\
 & & \label{4.14c} 
\end{eqnarray}
\end{mathletters}
Here the arguments $\vec{\kappa}=1-6$ of $v'_{\nu}{}^{(\alpha)}$
stand for $\vec{X}(\vec{\kappa})$, with
$\vec{X}(1)=(0,1,1)$, $\vec{X}(2)=(1,0,1)$, $\vec{X}(3)=(1,1,0)$,
$\vec{X}(4)=(0,-1,1)$, $\vec{X}(5)=(-1,0,1)$, $\vec{X}(6)=(-1,1,0)$
in units $a/2$, where $a$ is the cubic lattice constant.
Introducing the homogeneous strains
\begin{eqnarray}
   \lim_{\vec{k} \rightarrow 0} ik_{\mu} \langle u_{\nu}(\vec{k}) \rangle
   =\sqrt{mN} \epsilon_{\mu \nu},\;\;\; \nu=x,y,z,
\label{4.6}
\end{eqnarray}
we obtain
\begin{eqnarray}
  F_{QQT}^{(\alpha)}[\epsilon,\rho]/N&=&\rho_{\alpha}^2 [
  \Lambda^{(\alpha)}_{xx} (\epsilon_{xx}+\epsilon_{yy}) +
  \Lambda^{(\alpha)}_{zz} \epsilon_{zz}
   \nonumber \\
  & & + 2\Lambda^{(\alpha)}_{xy}\epsilon_{xy} +
  2\Lambda^{(\alpha)}_{xz}(\epsilon_{xz}+\epsilon_{yz}) ]. 
\label{4.7} 
\end{eqnarray}
(From the symmetry of the order parameters it follows that
$\Lambda^{(\alpha)}_{xx}=\Lambda^{(\alpha)}_{yy}$ and
$\Lambda^{(\alpha)}_{xz}=\Lambda^{(\alpha)}_{yz}$.)
It is convenient to work
in the system of axes which reflects the monoclinic symmetry,
Fig. 2. We therefore consider the coordinate system $x'y'z'$,
where $x'$ axis corresponds to $[1{\bar 1}0]$, $y'$ - to $[110]$
and $z'$ to $[001]$ directions of the cubic system. 
Notice that the new axes are obtained by the clockwise rotation
about the $z$ axis by $\pi/4$. Since $\Lambda_{\mu \nu}$ ($\mu,\nu=x,y,z$)
is a tensor of the second rank, we write $F_{QQT}[\epsilon,\rho]$
in the new coordinate system as
\begin{eqnarray}
  & & F^{(\alpha)}_{QQT}[\epsilon',\rho]/N=  \rho_{\alpha}^2  \nonumber \\
  & & \times [
  {\Lambda'}^{(\alpha)}_{xx} {\epsilon'}_{xx}+
  {\Lambda'}^{(\alpha)}_{yy} {\epsilon'}_{yy} +
  {\Lambda'}^{(\alpha)}_{zz} {\epsilon'}_{zz}+ 
  2{\Lambda'}^{(\alpha)}_{yz} {\epsilon'}_{yz}] .
\label{4.8} 
\end{eqnarray}
In the transformed coordinate system the elastic term of the
free energy reads
\begin{eqnarray}
  F_{TT}[\epsilon']/(V_c N)& & = \frac{1}{2}c_5
  ({\epsilon'}_{xx}^2+{\epsilon'}_{yy}^2)
  +\frac{1}{2}c_{11} {\epsilon'}_{zz}^2 \nonumber \\
  & &+c_{12}({\epsilon'}_{xx}+{\epsilon'}_{yy}){\epsilon'}_{zz} 
  +c_6 {\epsilon'}_{xx} {\epsilon'}_{yy} \nonumber \\  
  & &+(c_{11}-c_{12}) {\epsilon'}_{xy}^2+ 
  2c_{44} {\epsilon'}_{yz}^2 ,
\label{4.9} 
\end{eqnarray}
where we have introduced the notations 
$c_5=(c_{11}+c_{12})/2+c_{44}$, $c_6=(c_{11}+c_{12})/2-c_{44}$
and $c_{11}$, $c_{12}$, $c_{44}$ are the cubic elastic constants.
Minimizing $F_{QQT}+F_{TT}$ with respect to the strains ${\epsilon'}_{\mu \nu}$
for a given configuration with a fixed expectation value $\rho_{\alpha}$
we obtain:
\begin{mathletters}
\begin{eqnarray}
 {\epsilon'}_{xx}&=&-\frac{\rho_{\alpha}^2}{2c_{44} \triangle V_c}
 [{\Lambda'}{}_{xx}^{(\alpha)}(c_{11}c_5-c_{12}^2) 
  -{\Lambda'}{}_{yy}^{(\alpha)} (c_{11}c_6-c_{12}^2) 
  \nonumber \\
  & &-2{\Lambda'}{}_{zz}^{(\alpha)} c_{12}c_{44}] ,  \label{4.10a} \\ 
 {\epsilon'}_{yy}&=&-\frac{\rho_{\alpha}^2}{2c_{44} \triangle V_c}
 [{\Lambda'}{}_{yy}^{(\alpha)}(c_{11}c_5-c_{12}^2)
   -{\Lambda'}{}_{xx}^{(\alpha)}(c_{11}c_6-c_{12}^2) 
  \nonumber \\
 & &-2{\Lambda'}{}_{zz}^{(\alpha)} c_{12}c_{44}] ,  \label{4.10b} \\
  {\epsilon'}_{zz}&=&-\frac{\rho_{\alpha}^2}{\triangle V_c}
 [{\Lambda'}{}_{zz}^{(\alpha)} (c_{11}+c_{12})
 -c_{12}({\Lambda'}{}_{xx}^{(\alpha)}+{\Lambda'}{}_{yy}^{(\alpha)})] , 
 \nonumber \\
 & &  \label{4.10c} \\
 {\epsilon'}_{yz}&=&-\frac{\rho_{\alpha}^2}{2c_{44} V_c}
 {\Lambda'}{}_{yz}^{(\alpha)} ,      \label{4.10d} 
\end{eqnarray}
\end{mathletters}
where $\triangle=-2c_{12}^2+c_{11}(c_{11}+c_{12})$.
The shear distortion ${\epsilon'}_{yz}$ implies that
in the monoclinic phases the angle between the axes $y'_m$ and $z'_m$
attached to the crystal
deviates from $\pi/2$ by $\alpha \approx 2 {\epsilon'}_{yz}$.

We now present numerical results for the quadrupole order
in the $C2/c$ structure.
In the monoclinic system of axes $(x' y' z')$ the calculated
values are quoted in Table~IV.
(The shear angle $\alpha \approx -32 \cdot 10^{-4} \rho_1^2$.)
%
%   TABLE 4
%   -------
\begin{table} 
\caption{
 Calculated parameters $\Lambda'{}_{\mu \nu}^{(1)}$ and
 homogeneous strains $\epsilon'_{\mu \nu}$ for
 the quadrupolar ordering $Fm{\bar 3}m \rightarrow C2/c$;
 $\rho_1$ is the order parameter amplitude.
\label{tab4}     } 
 
 \begin{tabular}{c | c c c c } 
$\mu\; \nu$  & $xx$ & $yy$ & $zz$ & $yz$ \\
\tableline
$\Lambda'{}_{\mu \nu}^{(1)}$    &  155.7 K & 94.2 K & 311.1 K & 243.6 K  \\
$\left[ \epsilon'_{\mu \nu}/(\rho_1)^2 \right] \times 10^4$  
                        &  -2.08   & +1.95  &  -4.66  & -16.0 
 \end{tabular} 
\end{table} 
For calculations of $\epsilon'_{\mu \nu}$ we took the elastic
constants $c_{11}=10285$, $c_{12}=3969$ and $c_{44}=1188$ in units
K/{\AA}$^3$ from Ref.~\onlinecite{Mat3}.
Returning now to the original cubic system of axes $(x,y,z)$
we find
\begin{eqnarray}
& & \begin{array}{l l}
   {\epsilon}_{xx}={\epsilon}_{yy}=-0.07 \cdot 10^{-4} \rho_1^2, & 
   {\epsilon}_{zz}=-4.66 \cdot 10^{-4} \rho_1^2, \\
   {\epsilon}_{xy}=+2.02 \cdot 10^{-4} \rho_1^2, & 
   {\epsilon}_{xz}={\epsilon}_{yz}=-11.3 \cdot 10^{-4} \rho_1^2 .  
\end{array} \nonumber \\
& & \label{4.12a} 
\end{eqnarray}

For the condensation scheme (\ref{3.16b})
to the $C2/m$ quadrupole structure the calculated values
are quoted in Table~V (the shear angle 
$\alpha=-6.84 \cdot 10^{-4} \rho_2^2$).
%
%   TABLE 5
%   -------
\begin{table} 
\caption{
 Calculated parameters $\Lambda'{}_{\mu \nu}^{(2)}$ and
 homogeneous strains $\epsilon'_{\mu \nu}$ for
 the quadrupolar ordering $Fm{\bar 3}m \rightarrow C2/m$;
 $\rho_2$ is the order parameter amplitude.
\label{tab5}     } 
 
 \begin{tabular}{c | c c c c } 
$\mu\; \nu$  & $xx$ & $yy$ & $zz$ & $yz$ \\
\tableline
$\Lambda'{}_{\mu \nu}^{(2)}$    &  9.2 K & 488.9 K & 63.0 K & 52.3 K  \\
$\left[ \epsilon'_{\mu \nu}/(\rho_2)^2 \right] \times 10^4$  
                        &  +12.61   & -18.87  &  +1.46  & -3.42 
 \end{tabular} 
\end{table} 
In the initial cubic system of axes we find
\begin{eqnarray}
& & \begin{array}{l l}
   {\epsilon}_{xx}={\epsilon}_{yy}=-3.13 \cdot 10^{-4} \rho_2^2, & 
   {\epsilon}_{zz}=+1.46 \cdot 10^{-4} \rho_2^2, \\
   {\epsilon}_{xy}=-15.74 \cdot 10^{-4} \rho_2^2, & 
   {\epsilon}_{xz}={\epsilon}_{yz}=-2.42 \cdot 10^{-4} \rho_2^2 .  
\end{array} \nonumber \\
& & \label{4.16} 
\end{eqnarray}

We conclude that the two possibilities of quadrupole order
lead to completely different displacements in the monoclinic
phase.

\section {Conclusions} 
\label{sec:con} 

We present a microscopic model of quadrupole order in TmTe.
For a 4$f$ hole above $T_Q$ we have obtained the sequence
$\Gamma_7-\Gamma_8-\Gamma_6$ of the crystal electric field (CEF) 
energy spectrum with $\Gamma_7$ as ground state 
which is in agreement with results from
M\"{o}ssbauer spectroscopy \cite{Tri}
and ultrasonic velocity measurements. \cite{Kas}
The splitting of CEF levels is found to be small 
if only contributions from six Te
nearest neighbors of a Tm site are taken into account.

We have considered quadrupolar interactions between 4$f$
holes located on Tm sites.
On the basis of neutron diffraction experiments \cite{Lin2}
indicating 
that a single arm of $^*\vec{q}_L$ is responsible for the
quadrupole structure, 
we have studied the quadrupole interactions at the $L$ point
of the BZ. We have found that the quadrupole coupling
between 4$f$ holes becomes
attractive at the $L$ point thus driving a structural
phase transition with concomitant lowering of the crystal
symmetry. 
Starting with the mean-field Hamiltonian we have 
derived the Landau free energy, calculated
the transition temperature and found $T_c=4.9$~K.   
The overestimation of
the transition temperature is ascribed to 
 a screening effect from conduction electrons which
has not been considered in the present work.
The structure of TmTe below $T_Q$ is monoclinic (Fig.~2) but
there are still two possibilities for the quadrupole order parameter.
These quadrupole order parameters are expressed
in real space in
terms of $T_{2g}$ and $E_g$ components and visualized
in Fig.~1. The condensation of $\rho^F_1$, Eq.~(\ref{3.16a}),
leads to the $C2/c$ structure while the
condensation of $\rho^F_2$, Eq.~(\ref{3.16b}), leads to
the $C2/m$. Although both structures are monoclinic
their symmetries are different. Both of them result in the
domain variants which have been observed experimentally.
We conclude that on the basis of the present experimental data
and our theoretical studies it is impossible to
determine unambiguously the actual quadrupole order in TmTe.
We have shown that such discrimination could be done in respect
to lattice distortions which develop below $T_Q$.
Starting from the quadrupole-quadrupole interactions on a
deformable lattice, we have derived the couplings of the quadrupoles 
with the atomic lattice displacements.
We have calculated the corresponding lattice distortions
and suggest experiments which can be decisive
in determining which quadrupole order is realized in TmTe.

%------- ACKNOWLEDGMENTS ------------------------ 
\acknowledgments 
We thank J.M. Mignot and A. Gukasov for useful discussions.
This work has been financially supported by the Fonds voor
Wetenschappelijk Onderzoek, Vlaanderen.
 
\appendix 
%------- APPENDIX A ------------------------ 
\section{} 
\label{sec:apA}

Here we investigate transformational properties
of functions $S^{(1)}(\Omega)$ and $S^{(2)}(\Omega)$,
Eqs.~(\ref{3.12a},b) (see Fig.~1). Below we omit the argument $\Omega$
of the functions. We recall that the functions $S_{T_{2g}}^1$,
$S_{T_{2g}}^2$ and $S_{T_{2g}}^3$ are proportional to the
Cartesian components $yz$, $zx$ and $xy$, respectively.
Therefore, for the reflection $m$ through the plane
$[1{\bar 1}0]$
\begin{eqnarray}
& &m (S_{T_{2g}}^1 - S_{T_{2g}}^2) = -(S_{T_{2g}}^1 - S_{T_{2g}}^2) , \\
& &m (S_{T_{2g}}^1 + S_{T_{2g}}^2) = +(S_{T_{2g}}^1 + S_{T_{2g}}^2) , 
 \label{A.1}
\end{eqnarray}
while for the rotation $C_2$ by $\pi$ about the axis $[1{\bar 1}0]$
\begin{eqnarray}
& &C_2 (S_{T_{2g}}^1 - S_{T_{2g}}^2) = -(S_{T_{2g}}^1 - S_{T_{2g}}^2) , \\
& &C_2 (S_{T_{2g}}^1 + S_{T_{2g}}^2) = +(S_{T_{2g}}^1 + S_{T_{2g}}^2) . 
 \label{A.2}
\end{eqnarray}
Therefore,
\begin{eqnarray}
& &m S^{(1)} = -S^{(1)} ,   \\
& &C_2 S^{(1)} = -S^{(1)} , 
 \label{A.3}
\end{eqnarray}
while
\begin{eqnarray}
& &m S^{(2)} = S^{(2)} ,   \\
& &C_2 S^{(2)} = S^{(2)} . 
 \label{A.4}
\end{eqnarray}

%---------------- REFERENCES ------------------------------- 

%-------------- FIGURE CAPTIONS ------------------------------ 


\begin{references}
\bibitem{full}
The influence of the interstitial electron charge density
on CEF calculation will be discussed in a separate paper.
\end{references}
\end{document}